\documentclass[12pt]{revtex4}
\usepackage{amsmath,epsfig,amssymb}
\def\beq{\begin{equation}}
\def\eeq{\end{equation}}
\def\bea{\begin{align}}
\def\eea{\end{align}}

\newcommand{\g}{\gamma}

\renewcommand{\t}{\theta}

\renewcommand{\d}{\delta}

\begin{document}

\title[Power grids vulnerability: a complex network approach]{Power grids vulnerability:\\
a complex network approach}

\author{S.~Arianos$^{1,2}$, E.~Bompard$^1$, A.~Carbone$^2$ and F.~Xue$^1$}
\affiliation{$^1$ Dipartimento di Ingegneria Elettrica, Politecnico di Torino, Corso Duca degli Abruzzi 24, 10129 Torino, Italy}
\email{sergio.arianos@polito.it, ettore.bompard@polito.it, fei.xue@polito.it}

\affiliation{$^2$ Dipartimento di Fisica, Politecnico di Torino, Corso Duca degli Abruzzi 24, 10129 Torino, Italy}
\email{anna.carbone@polito.it}
%\date{\today}

\begin{abstract}
Power grids exhibit patterns of reaction to outages similar to complex networks. Blackout sequences follow power laws, as complex systems operating near a critical point.
Here, the tolerance of electric power grids to both accidental and malicious outages is analyzed in the framework of complex network theory. In particular, the quantity known as \emph{efficiency} is modified by introducing a new concept of \emph{distance} between nodes. As a result, a new parameter called \emph{net-ability} is proposed to evaluate the performance of power grids. A comparison between \emph{efficiency} and \emph{net-ability} is provided by estimating the \emph{vulnerability} of sample networks, in terms of both the metrics. 
\end{abstract}

\maketitle

{\bf Technological infrastructures are of vital importance for contemporary societies. As a consequence of the world wide growing interconnections, the security of networks such as world-wide-web, transport, power systems, is becoming a priority in the agenda of policy-makers, industrial and academic researchers. In recent years several blackouts occurring in USA and Europe have drawn a lot of attention to security problems in electric power transmission systems. In these scenarios, it is convenient to go beyond the traditional deterministic \emph{bottom-up} description in favor of a statistical \emph{top-down} approach.  Also the specific area of power systems has attracted the physicists community interested in the applications of complex network theory. In this paper, we investigate the topological structure and resilience of power grids by adopting a complex network description. We notice that the geodesic distance, used in complex network metrics, can be generalized to account for the flow capacity between nodes. Based on this new concept of distance, a metric called \emph{net-ability} is introduced to estimate the performance and resilience of power networks upon line removal.}

\section{Introduction}
Modern states and societies can only function if the necessary infrastructures are continuously available and fully operative. Critical infrastructures are organizations or facilities of key importance to public interest whose failure or impairment could result in detrimental supply shortages, substantial disturbance to public order or economic impact. The theory of complex networks is increasingly being exploited to tackle those sorts of issues. For a comprehensive review on complex networks we refer to \cite{boccaletti}. Examples of applications include facilities for electricity generation, transmission and distribution, oil and gas production, telecommunication, water supply, agriculture (food production and distribution), public health (hospital, ambulances), transportation systems (fuel supply, railways, airports, harbors), financial and security services \cite{albert1}-\cite{lawniczak}. Due to their importance, a crucial issue is learning how to improve the tolerance of critical infrastructure to failures and attacks.  A line of research investigating issues of flow and transportation in complex networks is under active development \cite{meloni}-\cite{danila}.
A major threat for the proper functioning of power networks is that of large blackouts that may involve big cities or even portions of states. Traditionally such occurrences were caused by accidental faults and thus were quite rare; however, in recent years power systems, as well as other critical infrastructures, have become a potential target for intentional attacks. The main difference is that malicious attacks may not be random but rather directed specifically to the most sensitive parts of the system, in terms of the impact they can cause. Thus, most of the applications of complex network concepts to power systems are aimed at understanding the behavior of power grids both in case of accidental failures and of malicious attacks \cite{crucitti3}-\cite{sole}. 

The tolerance of a network to failures is normally intended as the ability of the network to maintain its connectivity properties after the deletion of a fraction of its nodes or lines. In this way the problem can be mapped into a standard percolation problem, of the type extensively studied in statistical physics \cite{rosato}, \cite{rosas}. However, a pure connectivity approach, which may be suitable for instance in the case of the World Wide Web, does not seem to catch most of the crucial features of a power network. In general a power network can indeed undergo severe damages even without any inverse percolation taking place; on the contrary, it can happen that some less important nodes become isolated, thus changing the connectivity of the network, without strongly affecting its global performance.

A parameter introduced to evaluate the tolerance of complex networks to outages is the \emph{efficiency} \cite{latora1}. In the present paper we further develop this concept and  propose a new parameter to evaluate the performance of a power grid, which we name \emph{net-ability}. The new definition takes into account some peculiar features of electrical networks such as the flow limits and the power flow allocation through the network, due to the inherent physical laws.

The paper is structured as follows: in Section \ref{efvul} we review the definition of efficiency pointing out its meaning in relation to power grids. In Section \ref{netab} we introduce the definition of net-ability. In Section \ref{case} we provide examples of the application of net-ability to evaluate the static tolerance to outages of a few sample networks. Comparison between efficiency and net-ability is finally provided. General conclusions and comments are provided in Section \ref{concl}. Finally, in the Appendix we recall some basic notions about power systems analysis used in the paper.

\section{Efficiency and vulnerability} \label{efvul}
As a preliminary step, let us briefly recall the definition of geodesic distance commonly used in the literature on complex networks. Let us start considering an un-weighted graph: the number of lines in a path connecting nodes $i$  and $j$ is called the length of the path. A geodesic path (or shortest path) between $i$ and $j$ is the path connecting these nodes with minimum length. The length of the geodesic path is the \emph{geodesic distance} $d_{ij}$ between $i$  and $j$. If one is dealing with a weighted graph, the length of a path is the sum of the weights of the lines constituting that path.

The \emph{global efficiency} $E$ of a network was first introduced by Latora and Marchiori \cite{latora1} as follows:
\begin{equation}
E=\frac{1}{N(N-1)}\sum_{i\ne j}\frac{1}{d_{ij}}\;,
\label{topeff}
\end{equation}
where $N$ is the number of nodes of the network and $d_{ij}$ is the geodesic distance between the nodes $i$ and $j$ ; the sum is taken over all pairs of nodes of the network. The global efficiency is a measure of the performance of the network, under the assumption that the efficiency for sending information between two nodes $i$ and $j$ is proportional to the reciprocal of their distance.

In many networks it happens that some nodes and lines are more important than others. While naively one would say that the most important nodes are those with the highest degree \cite{footnote}, for large networks it is often non trivial to find out which are the components that actually are most critical for the performance of the network. Since the efficiency has been associated with the performance of the network, a natural way to find critical components of a network is by looking for the nodes or lines whose removal causes the biggest drops in efficiency. The \emph{vulnerability} $V_E(l)$ of a line $l$ can be defined as the drop in the performance when the line $l$ is removed from the network \cite{koganov}:
\begin{equation}
V_E(l)=\frac{E-E_l}{E}\;,
\label{vuln}
\end{equation}
where $E$ is the global efficiency of the network and $E_l$ is the global efficiency after the removal of the line $l$. When a node is removed, all the lines attached to the node are removed as well.

A definition of network vulnerability is the maximum vulnerability of all its nodes \cite{latora2}:
\begin{equation}
V=\max_{l}V(l)\;.
\end{equation}

The general definition of vulnerability, Eq.~(\ref{vuln}) as a drop in the efficiency can be usefully applied also to power networks. However, when applied to power grids, some problems arise with the definition of efficiency given by Eq.~(\ref{topeff}).

Specifically, the efficiency defined by Eq.~(\ref{topeff}) shows three main problems when applied to power grids.
\begin{itemize}
\item[1.] In electrical circuits power does not flow from a node $i$ to another node $j$ along a single specific path (for instance the geodesic path), but rather along all the paths connecting $i$ to $j$ according to the power flow; see the Appendix for a simplified method to solve the power flow equations. Therefore the classical idea of geodesic distance is not suited for power grids and a different concept of distance needs to be introduced.
\item[2.] In the Eq.~(\ref{topeff}) the sum is taken over all pair of nodes. However in electrical circuits power flows only from generation to load nodes, so only distances between generator-load pairs should be taken into account.
\item[3.] For each pair $(i,\,j)$ of generation and load nodes, the network has a different transfer capability $C_{ij}$ in transmitting power. Suppose we increase the power injection at node $i$ until the first line reaches its line flow limit: $C_{ij}$ is equal to the power injection in that moment.
\end{itemize}

\section{From efficiency to net-ability} \label{netab}
In the same spirit of the efficiency described in the previous section, the net-ability of a power transmission grid is defined as a measure of its performance under normal operating conditions. The function of a power transmission network is to transmit a time dependent amount of power from generation nodes to load nodes in the most convenient technical and economic way. The economic issues are related to transmission costs and economic efficiency (social surplus) of the underlying market, while the technical issues refer to losses, voltage drop and stability. The actual ability of a power transmission grid to perform properly depends on its topological structure and on the impedance and flow limits of its lines.

The concept of distance $d_{ij}$ may be explained as the difficulty to transfer the relevant quantity between the nodes $(i,\,j)$ of a network. Distance in general depends on the path that one follows and thus should be defined as a function of the characteristics of the lines along the path. The economic and technical difficulties for transmission of electrical power through a path depend on both the power flow through the lines and on their impedance: with the same impedance, higher power flow increases costs; with the same power flow, higher impedance increases costs. Consequently, the distance from node $i$ to node $j$ along path $k$ is related not only to  the impedance of each line of the path but also to the power flows through the lines of the path. As a result, we define the electrical distance as:
\begin{equation}
d_{ij}^k=\sum_{l\in k}f_k^lZ_l\;,
\label{eldist}
\end{equation}
where $f_k^l$ is the power transmission distribution factor of line $l$ in path $k$ and $Z_l$ its impedance (see Eq.~(\ref{ptdf}) in the Appendix for details) .

On account of the Eq.~(\ref{eldist}), we propose the following definition for the net-ability of a power transmission grid:
\begin{equation}
A=\frac{1}{N_GN_D}\sum_{i\in \mathcal{G}}\sum_{j\in \mathcal{D}}C_{ij}\sum_{k\in\mathcal{H}_{ij}}p_{ij}^k\frac{1}{d_{ij}^k}
 \label{neta1}
\end{equation}
where $\mathcal{G}$ and $\mathcal{D}$ are the sets of generator and load nodes respectively, while $\mathcal{H}_{ij}$ is the set of paths from generator $i$ to load $j$; likewise $N_G$ and $N_D$ are the total numbers of generators and loads respectively. Finally, $p_{ij}^k$ is the power share of path $k$ in transmitting power from $i$ to $j$.

Let us stress that the definition of distance given in the Eq.~(\ref{eldist}) is referred to a specific path. There is not the concept of geodesic distance or shortest path here, in principle all the paths are to be taken into account separately.

\begin{figure}
\includegraphics[width=16cm]{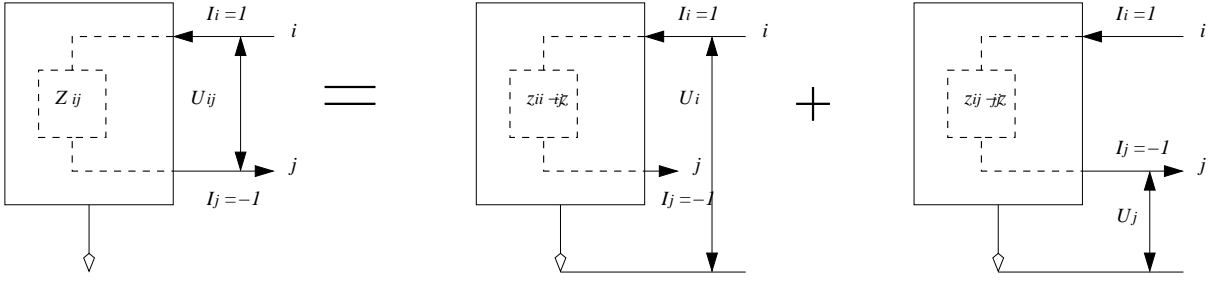}
\caption{The computation scheme of equivalent impedance} \label{zeqfig}
\end{figure}

Let us call $Z_{ij}$ the equivalent impedance of the circuit whose ends are node $i$ and node $j$; $U_{ij}$ is the voltage between $i$ and $j$ and $I_i$ is the current injected at node $i$ and extracted at node $j$ ($I_i=-I_j$). As shown in Fig.~\ref{zeqfig} the equivalent impedance is defined as
$$ Z_{ij}=\frac{U_{ij}}{I_i}\;. $$
Furthermore, let $I_i=1$, $I_j=-1$ and $I_h=0$ $\forall~h\ne i,\,j$ (meaning that a unit current is injected at node $i$ and extracted at $j$, while no current is extracted nor injected in other nodes), then the computation of equivalent impedance is sketched in Fig.~\ref{zeqfig} and amounts to
\begin{align}
Z_{ij}=\frac{U_{ij}}{I_i}=U_{ij} && \Rightarrow && Z_{ij}=U_i-U_j=(z_{ii}-z_{ij})-(z_{ij}-z_{jj})=z_{ii}-2z_{ij}+z_{jj}\;,
\end{align}
where $z_{ij}$ is the $i$-th, $j$-th element of the impedance matrix, see the Appendix.

In the following electrical networks are analyzed using a DC model. For a discussion of the reasons to choose a DC rather than an AC model, see the Appendix.
In a DC power flow model the distance $d_{ij}^k$ defined in Eq.~(\ref{eldist}) is equal to the variation of voltage angle between nodes $i$ and $j$, when 1 unit of active power is injected at $i$ and collected at $j$. Since in the DC power flow model the variation of voltage angle is considered as the equivalent DC voltage and the active power is considered as the current, $d_{ij}^k$ for any involved path between $i$ and $j$ is equal to the equivalent impedance of the circuit whose ends are $i$ and $j$.
\begin{equation}
d_{ij}^k=Z_{ij} \qquad \forall~k\;.
\label{eqimp}
\end{equation}
Again, the expression of the distance between $i$ and $j$ given in Eq.~(\ref{eqimp}) is not related to a specific path, since it takes into account all the existing paths between $i$ and $j$.

 \label{eltdist}

Substituting Eq.~(\ref{eqimp}) into Eq.~(\ref{neta1}) and keeping in mind that:
$$ \sum_{k\in\mathcal{H}_{ij}}p_{ij}^k=1\;, $$
we obtain:
\begin{equation}
A=\frac{1}{N_GN_D}\sum_{i\in \mathcal{G}}\sum_{j\in \mathcal{D}}\frac{C_{ij}}{Z_{ij}}\;.
\label{neta}
\end{equation}

In analogy with the expression of the vulnerability given by Eq.~(\ref{vuln}), we define the vulnerability of line $l$ as the net-ability drop caused by an outage (cut) of the line $l$:
\begin{equation} \label{wuln}
V_A(l)=\frac{A-A_l}{A}
\end{equation}

\section{Case study} \label{case}
In this section we use the two definitions, given by Eqs.~(\ref{vuln}) and (\ref{wuln}), to estimate the line vulnerability of IEEE sample networks \cite{IEEE}, made of 30 and 57 nodes. For each case, the results obtained from \emph{efficiency} and \emph{net-ability} are compared with the \emph{overload rate}, defined following \cite{Ejebe}. The overload in an electrical network is given by:

\begin{equation}
P=\sum_{l\in\mathcal{L}}\frac{|P_l|}{P_l^{lim}}\;,
\end{equation}
where $P_l$ is the power flow through line $l$ calculated by the DC load flow model (see the Appendix), $P_l^{lim}$ is the flow limit of line $l$ and the sum is taken over the set $\mathcal{L}$ of lines in the network. We are interested in the sensitivity of $P$ to line outages, therefore we call $P(l)$ the real power performance parameter of the network upon cutting line $l$. The \emph{overload rate} is then defined as:
\begin{equation}
W(l)=\frac{P(l)-P}{P}\;.
\label{overload}
\end{equation}

\begin{center}
\begin{figure}[hb]
\includegraphics[width=12cm]{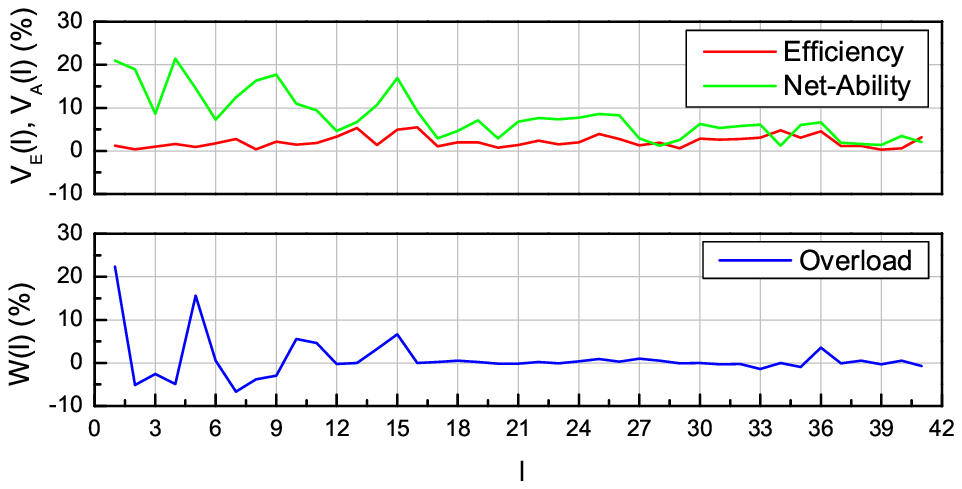}
\includegraphics[width=12cm]{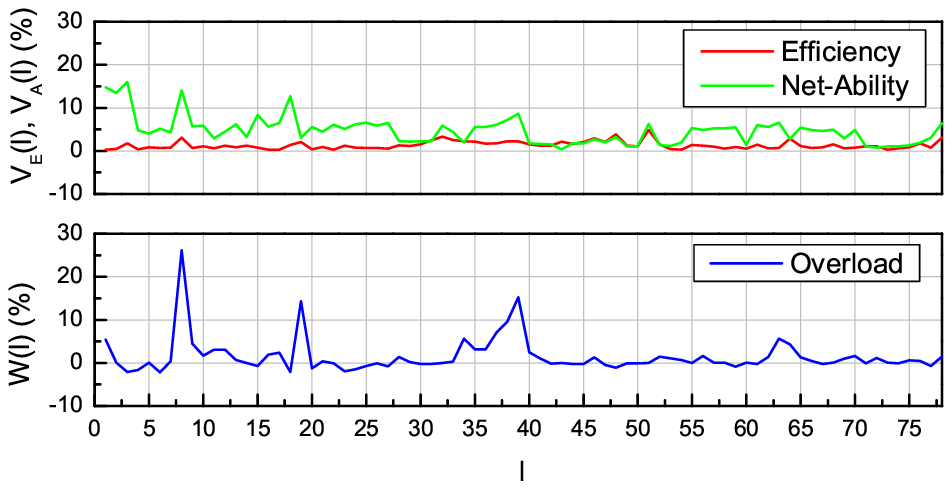}
\caption{Vulnerabilities and overload rate vs line removal for a 30 nodes, 41 lines IEEE test case (above); for a 57 nodes, 78 lines IEEE test case (below).}
\label{30S_tot}
\end{figure}
\end{center}

In figure  \ref{30S_tot}, we have plotted the vulnerability $V_E$ (Eq.~(\ref{vuln})), the vulnerability $V_A$ (Eq.~(\ref{wuln})) and the overload rate (Eq.~(\ref{overload})) versus line removal for the IEEE test cases with 30 and 57 nodes respectively. A few comments on the overload rate are appropriate. As we have shown, the overload rate is obtained by computing the power flow through each line of the network in the DC approximation. For a given network, the value of the DC power flow through each line is a non linear function of power injections (withdraws) at the generators (loads). On the other hand, those values are not taken into account in the definitions of efficiency and net-ability. Furthermore, in the IEEE test cases several generators produce pure reactive power, namely they are assigned a real power output equal to zero. On one side this means that these are not treated as generators in a DC flow model; on the other side these nodes are considered as generators both by the efficiency and net-ability algorithms. In order to overcome this limit, we have chosen to assign arbitrary values of active power output to the generators which are purely reactive. In table \ref{conversion}, we show explicitly these changes: in particular we keep the IEEE numeration for the generators, $Pg$ indicates the IEEE real power output, while $Pg'$ indicates the real power output assigned here.

\begin{center}
\begin{table}
\begin{tabular}{|c|c|c||c|c|c|}
\hline
Case30 & $Pg$ & $Pg'$ & Case57 & $Pg$ & $Pg'$ \\
\hline
node 1 & 260.2 & 260.2 & node 1 & 128.9 & 128.9 \\
\hline
node 2 & 40 & 40 & node 2 & 0 & 120 \\
\hline
node 5 & 0 & 210 & node 3 & 40 & 40 \\
\hline
node 8 & 0 & 130 & node 6 & 0 & 55 \\
\hline
node 11 & 0 & 95 & node 8 & 450 & 450 \\
\hline
node 13 & 0 & 78 & node 9 & 0 & 230 \\
\hline
& & & node 12 & 310 & 310 \\
\hline
\end{tabular}
\caption{Real power conversion for the IEEE 30- and 57-nodes generators.}
\label{conversion}
\end{table}
\end{center}

In conclusion, one can not expect a complete match between the results based on efficiency or net-ability and those based on the DC power flow. However, it appears from figure \ref{30S_tot} that in each of the sample cases under consideration the methods based on net-ability and overload rate computation can evidence a few highly critical lines; on the contrary, the plots obtained by the efficiency method are much smoother, without any sharp peak. In order to quantify this difference, in table \ref{variances} we report the variances $\sigma^2$ of the curves plotted in the figure \ref{30S_tot}. Moreover, the correlation coefficients $\rho$ between efficiency/overload, net-ability/overload and overload/overload are reported. We observe that the variances of the net-ability and overload curves are of the same order of magnitude, while those obtained from the efficiency curve are about one order of magnitude smaller. Likewise the correlation coefficients between net-ability/overload are significantly larger than those between efficiency/overload.
%\begin{center}
\begin{table}
\begin{tabular}{|c|c|c|c|}
\hline
& Efficiency & Net-Ability & Overload \\
\hline
$\sigma^2$ (30) & 1.94 & 30.55 & 24.05 \\
\hline
$\sigma^2$ (57) & 0.81 & 10.42 & 17.17 \\
\hline
$\rho$ (30) & 0.08 & 0.43 & 1 \\
\hline
$\rho$ (57) & 0.13 & 0.76 & 1 \\
\hline
\end{tabular}
\caption{Variances $\sigma^2$ of the curves of efficiency drop, net-ability drop and overload rate; correlation coefficients $\rho$ between efficiency/overload, net-ability/overload and overload/overload for the IEEE 30-nodes and 57-nodes test cases plotted in Figure 2.}
\label{variances}
\end{table}

\section{Conclusions} \label{concl}
In this paper, a new network metric called \emph{net-ability} is proposed, to evaluate the global performance of electric power grids. Our aim was to estimate the impact of line outages on the network performance in order to identify the most critical lines. In this respect we have analyzed sample networks taken from the IEEE database \cite{IEEE}. For each system, three different methods to evaluate the impact of line outages have been used: 1- the method based on efficiency; 2-  the new method based on net-ability; 3- the computation of line overloads by DC power flow. Since the latter is the approach which takes into account the specific details of power grids, it can be  considered as the reference method. From this point of view, the net-ability is capable of identifying some of the most critical lines. However, we stress how the line overloads depend non linearly upon the values of power injections/withdraws at the nodes. In real power networks such values are not constant, as the demand and production of electrical power vary considerably in time, for instance depending on the period of the year and the hours of the day. On the other hand, both the efficiency and the net-ability approach are based essentially on the topological features of the network and do not take into account the actual values of power injections and withdraws.

In order to check the validity of these topological approaches for real networks, one should compare their results with those obtained from overload computation, after performing a sort of time integration of the latter. At present such kind of time integration looks difficult to be implemented in an algorithm. However, it is actually performed by direct observation by those companies which are in charge of the management and control of power grids in each country and therefore have monitored each country network for years. We have investigated the data of the Italian power grid from Terna - Rete Elettrica Nazionale S.p.A. \cite{terna} in terms of net-ability, in order to find the most critical lines in the network. Although explicit results are confidential for obvious security reasons, we can say that a good match has been found between the results obtained by the net-ability algorithm and the experimental measurements collected by Terna.

\section*{Acknowledgements}
This work was supported by the Next Generation Infrastructures Foundation.

\appendix
\section{Linearized power systems models} \label{elett}
Here we provide a brief review of the main issues and tools of power system analysis used in this work; for a comprehensive treatment we refer to \cite{glover}.

A power transmission system can be schematically represented as a grid whose lines are electrical transmission lines, while nodes are the points where electrical power can be injected, withdrawn or redistributed. Accordingly, in a power grid one can distinguish three types of nodes: generation nodes (generators or power plants), load nodes (consumers), transmission nodes. Each line in a power network has its own maximum power flow capability, which is the maximum amount of power flow that the line can sustain.

Power transmission systems operate in a sinusoidal steady state. For a circuit made of $N+1$ buses operating in AC regime  the nodal equations are written as
\begin{equation} \label{matrix}
\left( \begin{array}{c}
\overline{I_1} \\ \overline{I_2} \\ \vdots \\ \overline{I_N}
\end{array} \right)~=~
\left( \begin{array}{cccc}
\overline{Y_{11}} & \overline{Y_{12}} & \cdots & \overline{Y_{1N}} \\
\overline{Y_{21}} & \overline{Y_{22}} & \cdots & \overline{Y_{2N}} \\
\vdots & \vdots & \ddots & \vdots \\
\overline{Y_{N1}} & \overline{Y_{N2}} & \cdots & \overline{Y_{NN}}
\end{array}\right)~
\left( \begin{array}{c}
\overline{U_1} \\ \overline{U_2} \\ \vdots \\ \overline{U_N}
\end{array} \right)
\end{equation}
where
$$ \overline{I_i}=I_ie^{\imath \t} \qquad \overline{Y_{ij}}=y_{ij}e^{\imath \g_{ij}} \qquad \overline{U_i}=U_ie^{\imath\d_i} $$
are complex quantities. In matrix notation Eq.~\ref{matrix} writes
\begin{equation}
\boldmath{I}=\boldmath{YU}
\end{equation}
where $\pmb{I}$ is the vector of  current sources, $\pmb{Y}$ is the line admittance matrix and $\pmb{U}$ is the vector of node voltages. Node $0$ is selected as the reference node (\emph{ground}), and node voltages $\overline{U_i}$ are defined with respect to node 0. The elements of the matrix $\pmb{Y}$ are formed as follows:
\begin{itemize}
\item diagonal elements $\overline{Y_{ii}}$: sum of the admittances of the lines connected to node $i$;
\item off-diagonal elements $\overline{Y_{ij}}$: minus the sum of the admittances of the lines connecting nodes $i$ and $j$.
\end{itemize}

The complex power $\overline{S_i}$ flowing through a node $i$ is defined as the product of the voltage $\overline{U_i}$ and the complex conjugate of the current $\overline{I_i}$:
\begin{equation}
\overline{S_i}=\overline{U_i}\overline{I_i}^*=\overline{U_i}\sum_{k=1}^N\overline{U_k}^*\overline{Y_{ki}}^*=P_i+\imath Q_i\;;
\end{equation}
$P$ and $Q$ are called \emph{real} and \emph{reactive} power respectively, see \cite{glover} for throughout explanation.

Since the quantities involved in a AC system show a sinusoidal behaviour, solving a full AC power flow model means that one has to solve a system of non linear equations, which is widely known to be a formidable task. The most common method to reduce the power flow problem to a set of linear equations is called the \emph{DC power flow}.

%As an example let us consider, within a bigger circuit, a single line connecting nodes $i$ and $j$.
The DC power flow approach is based on a number of approximations:
\begin{itemize}
\item reactive power balance equations are ignored;
\item line losses are ignored, that is the resistance of each line is set to zero, so only the reactance (imaginary part of the impedance) is considered: $y_{ii}=y_{ij}=y_{jj}=1/x_{ij}$ ($x_{ij}$ is the reactance of the line connecting $i$ to $j$), $\g_{ii}=\g_{jj}=-\frac{\pi}{2}$ and $\g_{ij}=\frac{\pi}{2}$;
\item all voltage magnitudes are identically set to one per unit, $U_i=1~~\forall\,i$;
\item all voltage angles are assumed to be small, $\d_i\to 0~~\forall\,i$.
\end{itemize}
Under these approximations, the power flow through line $l$, connecting nodes $i$ and $j$, is given by
\begin{equation}
f_l=P_i=-P_j=\frac{U_iU_j\sin(\d_i-\d_j)}{x_{ij}}=\frac{\d_i-\d_j}{x_{ij}}\;.
\end{equation}

In a general circuit made of $N$ nodes and $L$ lines, where $\overline{P}$ is the vector of real power injections, $\overline{\d}$ the vector of phase angles and $\overline{f}$ the vector of power flows we have
\begin{align}
P_i &=\sum_{l:i\to\forall j}f_l=\sum_{j=1}^N\frac{\d_i-\d_j}{x_{ij}} \\
\overline{P} &=B\overline{\d}  \label{pbd}
\end{align}
Where $B$ is the $N\times N$ admittance matrix:
\begin{align*}
B_{ij} &=-\frac{1}{x_{ij}} \qquad \textrm{for}\quad i\ne j \\
B_{ii} &=\sum_{j\ne i}\frac{1}{x_{ij}}
\end{align*}

In terms of the vector of power flows we have
\begin{equation}
\overline{f}=H\overline{\d}
\label{fhd}
\end{equation}
where $H$ is the $L\times N$ transmission matrix:
\begin{align*}
H_{li} &=-H_{lj}=\frac{1}{x_{ij}} \\
H_{lk} &=0 \qquad\forall\,k\ne i,j
\end{align*}

The admittance matrix $B$ is singular since the sum of the  elements of each row is equal to zero: $\sum_{i=1}^NB_{ij}=0$ $\forall\,j$. This means that the total injection power is equal to zero:
$$ \sum_{i=1}^NP_i=0 \quad \Rightarrow \quad P_i=-\sum_{j\ne i}P_j $$
To avoid this redundancy a \emph{slack node}, for instance node $N$, is chosen and set $\d_N=0$. Thus one can eliminate the corresponding terms in power vectors and matrices without losing information. In this spirit, $B'$ and $H'$ are sub-matrices, obtained respectively from $B$ and $H$ by deleting the row and column (only the column in case of $H$) corresponding to the slack node $N$, while $\overline{\d}'$ and $\overline{P}'$ are respectively the vector of phase angles and vector of node power injections without the slack node $N$.

The matrix $B'$ can be inverted and thus one can rewrite Eqs.~(\ref{pbd}) and (\ref{fhd}) in terms of the modified vectors and matrices as
\begin{align}
\overline{\d}' &=B'^{-1}\overline{P}' \\
\overline{f} &=H'B'^{-1}\overline{P}'=A\overline{P}' \label{ptdf}
\end{align}
The Power Transmission Distribution Factors (PTDF) of the circuit are the entries of the matrix $A$ in Eq.~(\ref{ptdf}).


\begin{thebibliography}{10}
\bibitem{footnote} The degree of a node in an undirected graph is the number of lines attached to that node.
\end{thebibliography}
\end{document}